\documentclass[conference]{IEEEtran}

\usepackage{graphicx}
\usepackage{tabularx}

\usepackage[T1]{fontenc}
\usepackage[utf8]{inputenc}

\usepackage{todonotes}
\usepackage{csquotes}
\usepackage{nicefrac}
\usepackage{hyperref}
\usepackage{amssymb}
\usepackage{amsthm}
\usepackage{tikz}
\usetikzlibrary{shapes.geometric, positioning, quantikz}

\usepackage[font=footnotesize]{subcaption}
\usepackage{yquant}
\useyquantlanguage{groups}
\usepackage{pgfplots}

\newtheorem{example}{Example}

\usepackage[style=ieee, maxnames=4, minnames=1, date=year, doi=false,isbn=false,backend=biber]{biblatex}
\AtEveryBibitem{
  \clearname{editor}%
  \clearfield{series}%
  \clearfield{isbn}%
  \clearfield{issn}%
  \clearfield{volume}%
  \clearfield{number}%
  \clearfield{pages}%
 \clearfield{url}%
  \clearfield{doi}%
  \clearfield{urldate}%
}

\useyquantlanguage{groups}
\usetikzlibrary{fit}

\DeclareFieldFormat
  [misc, book]
  {title}{\mkbibquote{#1}}

\usepackage{flushend}

\begin{document}
\title{Towards \\Application-Aware Quantum Circuit Compilation}
\def\authorwidth{8mm}

\author{
	 \IEEEauthorblockN{Nils Quetschlich\IEEEauthorrefmark{1}\hspace*{\authorwidth} 
  Florian J. Kiwit\IEEEauthorrefmark{2}\IEEEauthorrefmark{3}\hspace*{\authorwidth}
  Maximilian A. Wolf\IEEEauthorrefmark{2}\IEEEauthorrefmark{3}\hspace*{\authorwidth}
  Carlos A. Riofrío\IEEEauthorrefmark{2}\hspace*{\authorwidth}\\
  Lukas Burgholzer\IEEEauthorrefmark{1}\hspace*{\authorwidth}
  Andre Luckow\IEEEauthorrefmark{2}\IEEEauthorrefmark{3}\hspace*{\authorwidth}
  Robert Wille\IEEEauthorrefmark{1}\IEEEauthorrefmark{4}}
  \IEEEauthorblockA{\IEEEauthorrefmark{1}Chair for Design Automation, Technical University of Munich, Germany}
  \IEEEauthorblockA{\IEEEauthorrefmark{2}BMW Group, Munich, Germany}
   \IEEEauthorblockA{\IEEEauthorrefmark{3}Ludwig Maximilian University, Munich, Germany}
  \IEEEauthorblockA{\IEEEauthorrefmark{4}Software Competence Center Hagenberg GmbH (SCCH), Austria}
  \IEEEauthorblockA{\{\href{mailto:nils.quetschlich@tum.de}{nils.quetschlich}, \href{mailto:lukas.burgholzer@tum.de}{lukas.burgholzer}, \href{mailto:robert.wille@tum.de}{robert.wille}\}@tum.de \\
          \{\href{mailto:florian.kiwit@bmw.de}{florian.kiwit}, \href{mailto:maximilian.ww.wolf@bmw.de}{maximilian.ww.wolf}\}@bmw.de  \hspace*{\authorwidth} 
          \{\href{mailto:carlos.riofrio@bmwgroup.com}{carlos.riofrio}, \href{mailto:andre.luckow@bmwgroup.com}{andre.luckow}\}@bmwgroup.com\\
	 \url{https://www.cda.cit.tum.de/research/quantum} 
	 }
	 \vspace{-8mm}
}

\maketitle

\begin{abstract}
Quantum computing has made tremendous improvements in both software and hardware that have sparked interest in academia and industry to realize quantum computing applications.
To this end, several steps are necessary:
The underlying problem must be encoded in a quantum circuit, a suitable device must be selected to execute it, and it must be \emph{compiled} accordingly.
This compilation step has a significant influence on the quality of the resulting solution.
However, current \mbox{state-of-the-art} compilation tools treat the quantum circuit as a sequence of instructions without considering the actual application it realizes---wasting a yet untapped potential to increase the solution quality.
In this work, a different approach is explored that explicitly incorporates the application considered and aims to optimize its solution quality during compilation.
Initial results show the benefits of this approach: For an industry-inspired application of a quantum generative model, the proposed approach outperformed Qiskit's \mbox{most-optimized} compilation scheme and led to better solution quality.
Therefore, this work presents a first step towards \emph{\mbox{application-aware}} compilation.
\end{abstract}

\section{Introduction}
Quantum computing has made tremendous progress in both hardware and software development in recent years.
This has sparked interest not only in academia but also in industry with an increasing number of quantum computing applications being explored~\cite{bayerstadler_industry_2021}.
Realizing such applications using quantum computing has gained significant attention in various domains such as finance~\cite{stamatopoulosOptionPricingUsing2020}, logistics~\cite{harwoodFormulatingSolvingRouting2021}, and chemistry~\cite{chemapplication}.
Usually, these approaches follow a similar \emph{workflow} consisting of multiple steps (such as, e.g., described in~\cite{quetschlich2023mqtproblemsolver}).

Starting with the problem itself, a respective quantum algorithm generally capable of solving it must be selected together with either a quantum simulator or an actual device.
Then, the problem must be \emph{encoded} in a quantum circuit based on the selected algorithm.
This circuit must then be \emph{compiled} into an executable quantum circuit for the chosen device (if a \mbox{noise-free} simulator is selected, this step can be skipped). 
The compiled quantum circuit can be \emph{executed} afterwards---resulting in a histogram of measurement results that must be \emph{decoded} to extract the actual solution.
Workflows like this help end users realize their applications and can also be extended to cover further aspects, e.g., resource estimation~\cite{quetschlich2024utilizing}.

In the following, the focus is on the compilation step, which greatly affects the quality of the solution of the considered application and, therefore, can make the difference between a successful execution and obtaining completely random results.
To achieve the best possible solution, current compilers usually offer different levels of optimization as a \mbox{trade-off} between the compilation time and \emph{quality} of the compiled circuit.
The quality of a circuit is typically defined based on certain \emph{figures of merit} that act as a \emph{proxy} for the true solution quality of the considered application.
However, currently established figures of merit do not consider the actual application realized by the quantum circuit being compiled in their evaluation of the circuit's quality.

\begin{figure*}[t]
\centering
\resizebox{0.99\linewidth}{!}{
\begin{tikzpicture}
\begin{yquantgroup}
\registers{
qubit {} q[4];
}
\circuit{
        init {$q_0$} q[0];
        init {$q_1$} q[1];
        init {$q_2$} q[2];
        init {$q_3$} q[3];
        box {$R_Z(\pi)$} q[0];
        box {$H$\vphantom{$R_Z(\pi)$}} q[0-1];
        cnot q[2] | q[0];
        cnot q[3] | q[1];
        box {$R_Z$} q;
        cnot q[1] | q[0];
        cnot q[3] | q[2];
}
\equals[$\Longrightarrow$]
\circuit{
        [name=start]
        box {$R_Z(\pi)$} q[0];
        align q[0], q[1];
        [name=end]
        box {$R_Z(\frac{\pi}{2})$} q[0];
        box {$R_Z(\frac{\pi}{2})$} q[1];
        box {$SX$\vphantom{$R_Z(\frac{\pi}{2})$}} q[0-1];
        box {$R_Z(\frac{\pi}{2})$} q[0-1];
        cnot q[2] | q[0];
        cnot q[3] | q[1];
        box {$R_Z$} q;
        cnot q[1] | q[0];
        cnot q[3] | q[2];
\node[draw, dashed, blue, fit=(start)(end), rounded corners, inner xsep=1pt,inner xsep=1.5pt,inner ysep=1.5pt, line width=0.5mm] {};
}
\equals[$\Longrightarrow$]
\circuit{
        init {$q_3\mapsto Q_1$} q[0];
        init {$q_1\mapsto Q_2$} q[1];
        init {$q_0\mapsto Q_3$} q[2];
        init {$q_2\mapsto Q_4$} q[3];
        box {$R_Z(\frac{\pi}{2})$} q[1];
        box {$R_Z(\frac{3\pi}{2})$} q[2];
        box {$SX$\vphantom{$R_Z(\frac{\pi}{2})$}} q[1-2];
        box {$R_Z(\frac{\pi}{2})$} q[1-2];
        cnot q[0] | q[1];
        cnot q[3] | q[2];
        box {$R_Z$} q;
        [name=start2]
        cnot q[2] | q[0];
        cnot q[0] | q[2];
        [name=end2]
        cnot q[2] | q[0];
        cnot q[1] | q[0];
        cnot q[2] | q[3];
\node[draw, blue, dashed, fit=(start2)(end2), inner ysep=24pt, rounded corners, yshift=0.7cm, line width=0.5mm] {};
}
\end{yquantgroup}
    \def\xoffset{13.4}
    \def\yoffset{1.0}
    \node[] (abc) at (7.0,0+\yoffset) {
\begin{yquantgroup}
\registers{
qubit {} q[1];
}
\circuit{
h q[0];
}
\equals
\circuit{
        box {$R_Z(\frac{\pi}{2})$} q[0];
        box {$SX$\vphantom{$R_Z(\frac{\pi}{2})$}} q[0];
        box {$R_Z(\frac{\pi}{2})$} q[0];
}
\end{yquantgroup}
    };
    \node[shape=circle,draw=black,minimum size=0.3cm, inner sep=1pt] (0) at (\xoffset,0+\yoffset) {$Q_0$};
    \node[shape=circle,draw=black,minimum size=0.3cm, inner sep=1pt] (1) at (0.7+\xoffset,0+\yoffset) {$Q_1$};
    \node[shape=circle,draw=black, minimum size=0.3cm, inner sep=1pt] (2) at (1.4+\xoffset,0+\yoffset) {$Q_2$};
    \node[shape=circle,draw=black,minimum size=0.3cm, inner sep=1pt] (3) at (0.7+\xoffset,-0.7+\yoffset) {$Q_3$};
    \node[shape=circle,draw=black,minimum size=0.3cm, inner sep=1pt] (4) at (0.7+\xoffset, -1.4+\yoffset) {$Q_4$};
    \path [line width =1mm] (1) edge node[] {}(2);   
    \path [line width =1mm] (1) edge node[] {}(3);   
    \path [line width =1mm] (3) edge node[] {}(4);   
    \path [line width =1mm] (0) edge node[] {}(1);    
    \node [above of=1, yshift=-3mm] {ibmq\_quito};
    \path(4.south)+(0,-0.1) edge (\xoffset+0.7, \yoffset-2.1);  
    \path(5.55,\yoffset-0.1) edge (5.55, \yoffset-2.1);    
\end{tikzpicture}
}		
\subfloat[Quantum circuit.\label{fig:original_qc} ]{\hspace{.18\linewidth}}
\hspace{0mm}
\subfloat[Synthesized circuit.\label{fig:syn_qc}]{\hspace{.4\linewidth}}
\hspace{0mm}
\subfloat[Mapped circuit.\label{fig:map_qc}]{\hspace{.4\linewidth}}
\caption{Quantum circuit compilation flow.}
\vspace{-5mm}
\label{fig:compilation}
\end{figure*}
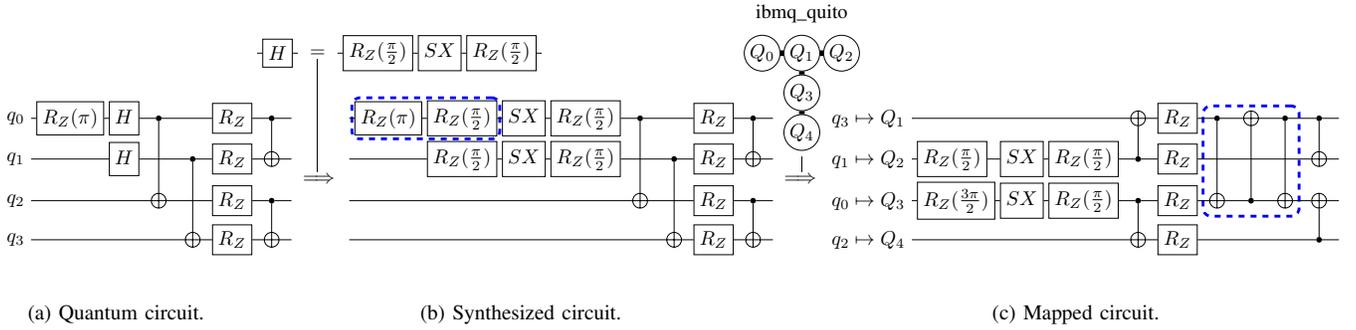

In this work, a different approach to compilation is explored.
Instead of compiling for a figure of merit that acts as a proxy for the resulting solution quality of a considered application, the solution quality \emph{itself} is used to guide the compilation.
For that, two steps are required:
Firstly, an \emph{\mbox{application-aware}} figure of merit must be defined that, obviously, must depend on the application considered.
It relates the compiled circuit to the quality of the solution of the application it realizes.
Second, a compilation environment is necessary that allows one to consider this customizable figure of merit. 

To explore the benefits of the proposed approach, it has been evaluated for an industry-inspired application of a \emph{quantum generative model}.
These generative learning applications are experiencing rapid growth and attracting widespread interest with applications extending from anomaly detection to text and image generation, as well as speech and video synthesis. 
However, generative learning still faces practical challenges, such as the need for large datasets, extensive hyperparameter tuning, long convergence times, and high computational resource requirements. 
\emph{Quantum generative modeling} might offer an opportunity to overcome these challenges and improve its capabilities by providing faster training, increased expressive power, improved inference, and reduced energy footprint~\cite{sim2019expressibility, du2020expressive,Villalonga_2020}. 
Furthermore, quantum systems are inherently probabilistic, which makes them a natural fit for modeling probability distributions.

\vspace{1cm}

To realize the proposed compilation scheme, an \mbox{application-aware} figure of merit is derived and used during compilation based on the MQT~Predictor framework~\cite{quetschlich2023mqtpredictor}---an open-source compilation framework that utilizes artificial intelligence to determine how to best compile a given quantum circuit for a customizable figure of merit.
Experimental evaluations for all considered application instances show that the proposed approach results in a higher solution quality compared to the \mbox{state-of-the-art} compiler Qiskit~\cite{qiskit} using its \mbox{most-optimized} compilation scheme.
Therefore, this work constitutes a first step towards \emph{application-aware} compilation.

The remainder of this work is structured as follows: 
In \autoref{sec:background}, the necessary background on quantum circuit compilation is provided along with a discussion of related work.
Then, \autoref{sec:motivation} motivates the proposed approach, which is described in more detail based on an exemplary application in \autoref{sec:proposed} and evaluated in \autoref{sec:eval}.
Lastly, the proposed approach is discussed in \autoref{sec:discussion} and concluded in \autoref{sec:conclusion}.

\section{Background}\label{sec:background}
This section describes how a given quantum circuit encoding any kind of application is compiled to become executable on a chosen device and reviews the related work for quantum circuit compilation.

\subsection{Quantum Circuit Compilation}
Starting with the quantum circuit that could contain \emph{any} quantum gate with no restrictions on the type and number of qubits used, an equivalent circuit shall be derived by \emph{compiling} it such that it adheres to the constraints induced by the chosen device.
To this end, different \emph{compilation passes} are used, namely \emph{synthesis}, \emph{optimization}, and \emph{mapping} as visualized in~\autoref{fig:compilation}.

\begin{example}\label{ex:qc}
    Assume that the application considered is encoded as the quantum circuit shown in \autoref{fig:original_qc}.
    This circuit shall now be compiled so that it becomes executable on the ibmq\_quito device with five qubits whose qubit connectivity is visualized above the arrow between \autoref{fig:syn_qc} and \autoref{fig:map_qc}.
\end{example}

Synthesis passes are run to ensure that all the quantum gates present in the circuit are translated into executable gate types---constituting the \mbox{so-called} \emph{native gateset} of a device.

\begin{example}\label{ex:syn}
    For the chosen ibmq\_quito device, the native gateset comprises the gates: \{$R_Z$, $SX$, $CX$, $X$, $ID$\}.
    However, the $H$ gates present in the circuit shown in \autoref{fig:original_qc} are not included and therefore must be \emph{synthesized} to a sequence of native gates---in this case as a sequence of $RZ$, $SX$, $RZ$ gates shown in \autoref{fig:syn_qc}.
\end{example}

\emph{Optimization} passes are typically run before and after synthesis passes to enable efficient synthesis and reduce the overhead introduced by translating to native gates.

\begin{example}
    The synthesized quantum circuit from \autoref{fig:syn_qc} can be optimized by merging the subsequent $R_Z$ gates (marked in dashed blue) into one gate by adding their angles.
\end{example}

Lastly, \emph{mapping} passes are run to assign each \emph{logical} qubit of the quantum circuit to a \emph{physical} qubit of the chosen device.
Often it is not possible to find a \emph{layout} such that the entire circuit adheres immediately to the device's connectivity. 
Typically, this is solved by inserting $SWAP$ gates into the circuit to dynamically permute the qubit arrangement on the device---a procedure called \emph{routing} that can lead to significant overhead in terms of additional quantum gates.
To counteract this potential increase in the number of gates, optimization passes are run again on the resulting circuit to reduce its size.
This finally results in a fully executable quantum circuit for the chosen device that represents the same functionality as the initial circuit, but usually has a very different structure.

\begin{example}
    For the circuit shown in \autoref{fig:syn_qc}, there is no layout that immediately satisfies the connectivity of the ibmq\_quito device which is shown above the arrow between \autoref{fig:syn_qc} and \autoref{fig:map_qc}.
    Therefore, one $SWAP$ gate is added, which has to be synthesized into native gates again---resulting in the now fully executable quantum circuit shown in \autoref{fig:map_qc} with the synthesized $SWAP$ gate highlighted in dashed blue.
\end{example}

\subsection{Related Work}
Since quantum circuit compilation is a very active field of research, various compilation methods have been proposed for synthesis~\cite{gilesExactSynthesisMultiqubit2013, amyMeetinthemiddleAlgorithmFast2013, millerElementaryQuantumGate2011, zulehnerOnepassDesignReversible2018, niemann2020advancedexactsynt, pehamDepthoptimalSynthesisClifford2023, bqskit_opt_syn}, optimization~\cite{bqskit_opt_syn, hattoriQuantumCircuitOptimization2018, vidalUniversalQuantumCircuit2004, itokoQuantumCircuitCompilers2019, maslovQuantumCircuitSimplification2008, niu2023powerful}, and mapping~\cite{lin_layout_synthesis, zulehnerEfficientMethodologyMapping2019,
matsuoEfficientMethodQuantum2019,
bochen_mapping,
willeMappingQuantumCircuits2019,
liTacklingQubitMapping2019,
pehamOptimalSubarchitecturesQuantum2023,
hillmichExploitingQuantumTeleportation2021,
zulehnerCompilingSUQuantum2019, bqskit_mapping, tket_mapping, schmidHybridCircuitMapping2024, MQT_QMAP}.
Over the years, these methods have been adopted and incorporated by many quantum \emph{Software Development Kits}~(SDKs) to create automated, powerful, and easily accessible compilation flows.
Prominent representatives from industry of such SDKs are IBM's Qiskit~\cite{qiskit}, Quantinuum's TKET~\cite{tket}, and Xanadu's~PennyLane~\cite{bergholm2022pennylane}.
Furthermore, various academic compilation software tools have matured to become more and more comprehensive, such as the Berkeley Quantum Synthesis Toolkit~\cite{bqskit}, and the Munich Quantum Toolkit~(MQT)~\cite{willeMQTHandbookSummary2024}~(e.g., MQT~QMAP~\cite{MQT_QMAP}).

All of the mentioned SDKs are capable of compiling a given quantum circuit for a chosen device and usually offer \mbox{pre-configured} sequences of compilation passes with an adjustable degree of optimization---constituting a \mbox{trade-off} between the time to create the compiled circuit and its quality. 
As a rule of thumb, the highest degree is used to derive the best possible compiled quantum circuit---however, usually no clear optimization criterion is given and it remains vague what the SDK is actually optimizing for.
Additionally, these SDKs try to address the needs of many and therefore follow a \enquote{one-fits-all} approach---aiming for a good \emph{average} performance across diverse quantum circuits and applications realized by them.
However, when considered a \emph{specific} application, this can lead to an untapped potential for the quality of its solution.

\section{Motivation}\label{sec:motivation}

Due to the variety of different available compilers and proposed compilation schemes, the respectively compiled quantum circuits also significantly differ, e.g., in terms of the number of gates, the mapping of logical to physical qubits, and the induced overhead by adhering to a device's restricted connectivity.
Although all compiled quantum circuits represent the same functionality and should lead to the same solutions to the application considered, this is not the case in reality due to the imperfection of the currently available \emph{Noisy Intermediate Scale Quantum}~(NISQ) computers~\cite{Preskill2018quantumcomputingin}.

This raises the question of how to evaluate the quality of a compilation scheme.
To this end, different metrics acting as \emph{figures of merit} have been proposed to compare compilation schemes.
Those figures of merit try to resemble how well a compiled circuit can be executed on a quantum computer and, by that, act as a \emph{proxy} for the solution quality of the considered application.
These figures of merit can be grouped into three categories with an increasing effort to calculate but also higher resemblance of the actual execution:

\begin{enumerate}
    \item Figures of merit based on characteristics of the compiled quantum circuit: This includes efficiently computable metrics such as the depth of the compiled circuit or its gate count---frequently limited to the number of \mbox{two-qubit} gates because, for most current architectures, their execution error rate is significantly higher than that of \mbox{single-qubit} gates.
    \item Figures of merit based on the characteristics of the compiled quantum circuit \emph{and} quantum device: Here, information such as specific gate error rates and qubit decoherence times is also considered, e.g., in the \emph{expected fidelity} used in \cite{quetschlich2023mqtpredictor} and the \emph{estimated success probability} described in \cite{dangwal2023clifford}.
    \vspace{1cm}
    \item Figures of merit based on the comparison of (noiseless) simulations and actual execution results of the compiled quantum circuit: When executed both on a noiseless simulator and on an actual quantum device, the overlap in the resulting histograms can be compared, such as, e.g., in \cite{dangwal2023clifford, salm2022prediction}. The compilation scheme leading to the largest overlap is selected as the most promising one.
\end{enumerate}

\begin{example}\label{ex:figure of merit}
To make those figures of merit more accessible, the compiled quantum circuit from \autoref{fig:map_qc} has been evaluated based on representatives of all three categories:
\begin{enumerate}
    \item Two-Qubit Gate Count: $7$
    \item Expected Fidelity~\cite{quetschlich2023mqtpredictor}: $0.5814$ 
    \item Histogram Intersection~\cite{salm2022prediction}: $0.6655$ 
\end{enumerate}
\end{example}

Although these figures of merit cover different aspects of the compiled circuit and its execution, they are used for specific purposes and neglect the end users' perspective, which prioritizes the solution quality of the considered application above all else.

Plain gate counts are typically used to assess single compilation passes, such as mapping, where these numbers directly reflect the passes' efficacy. %
However, they do not incorporate \mbox{hardware-induced} errors such as gate execution, measurement errors, and decoherence times---making them inadequate for assessing entire compilation flows. 
Integrating hardware characteristics through noise models derived from calibration data provides a rough estimate of circuit execution quality. 
Yet, these models often inadequately represent the actual device due to outdated or incomplete data as shown, e.g., in~\cite{dangwal2023clifford}. 
Comparing the histogram of noiseless simulations with the ones from actual executions of compiled circuits works well for \mbox{non-variational} quantum algorithms but may not be optimal for variational algorithms like QAOA~\cite{qaoa} (which is considered one of the most promising algorithms in the current NISQ era~\cite{Shaydulin_2019}), where noise could potentially aid optimization by preventing convergence to local minima, similar to \emph{gradient descent optimization algorithms}~\cite{ruder2017overview} used in classical machine learning.

Furthermore, all figures of merit currently used in compilation treat the \mbox{to-be-compiled} quantum circuit as a sequence of instructions without considering the actual application it realizes---wasting a yet untapped potential to increase the solution quality, which is exactly what counts for the potential end users.
This is especially unfavorable since the compilation itself can make the difference between a successful execution and obtaining completely random results.
Thus, compiling a given quantum circuit in the most promising fashion constitutes an important challenge in the development of quantum computing applications.

In this work, a different approach is explored: instead of defining a \emph{proxy} for the expected solution quality of a considered application and using this as the figure of merit during compilation, the solution quality of the application \emph{itself} is used---offering the potential to further increase it for a considered application.

\section{Proposed Approach}\label{sec:proposed}
\begin{figure*}[t]
    \centering
    \includegraphics[width=0.8\linewidth]{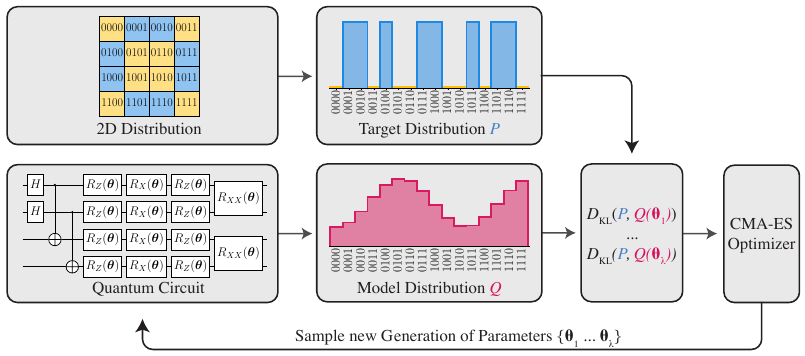}
    \caption{The QCBM fits the model distribution $Q$, generated by a parameterized quantum circuit, to the target distribution $P$, given by the dataset. For a population of size $\lambda$, the fitness of each candidate solution $\boldsymbol{\theta}_\mathrm{i}$ is evaluated. The \mbox{CMA-ES} optimizer samples a new population of candidate solutions, based on the \mbox{best-performing} candidate solutions of the previous generation.}
    \label{fig:application}
\end{figure*}

Realizing such an \emph{application-aware} compilation scheme for a considered application requires two steps: An \mbox{application-aware} figure of merit must be defined and a compilation environment must be set up that supports customizable figures of merit.
In the following, these steps are described in detail.
To this end, a running example of an \mbox{industry-inspired} application is used for illustrative purposes, which is introduced first.

\subsection{Quantum Generative Modeling}
Generative learning is experiencing rapid growth and is attracting widespread interest across industries, with applications ranging from anomaly detection to text and image generation, as well as speech and video synthesis. 
Ultimately, the objective of training a generative model is to express the underlying distribution of a dataset using a machine learning model. 
In \emph{quantum generative learning}, this model is represented by a parameterized quantum circuit~\cite{Benedetti_2019_2}.

\begin{example}\label{ex:QCBM}
A \emph{Quantum Circuit Born Machine}~(QCBM)~\cite{e20080583,Benedetti_2019,kiwit2023applicationoriented,kiwit2024benchmarking} is a quantum generative model that is trained as depicted in~\autoref{fig:application}.
To learn the target distribution---in this case, the distribution of a \mbox{two-dimensional} dataset that resembles the letter X (indicated by the orange tiles)---the parameter values of the shown parameterized quantum circuit are adjusted so that the resulting model distribution is as close as possible to the target distribution.
\end{example}

\subsection{Application-Aware Figure of Merit}
A figure of merit must be defined that takes into account the actual application.
For MaxCut problems, it might be the size of the cut.
For ground state estimation, it might be the current estimate for the ground state energy.
For the traveling salesperson problem, it might be the cost of the determined route.
Hence, such \emph{application-aware} figures of merit might look quite different depending on the considered application.
 
Furthermore, variational algorithms are executed in a hybrid \mbox{classical-quantum} fashion in which the parameter values of the underlying parameterized quantum circuit are optimized using a classical optimizer.
Those optimizers usually already optimize for an \mbox{application-aware} figure of merit that defines the solution quality of the considered application.
Therefore, the same metric should also be used during the compilation---aiming for the best overall resulting solution quality.

\begin{example}
During the QCBM training described in \autoref{ex:QCBM}, the deviation of the model distribution $Q$ to the target distribution $P$ shall be minimized.
This difference is determined using the \mbox{Kullback-Leibler}~(KL) divergence, 
$$
D_{\mathrm{KL}}(P \| Q)=\sum_{x \in \mathcal{X}} P(x) \log \left(\frac{P(x)}{Q(x)}\right).
$$
Whenever the KL~divergence reaches its minimum, the training is completed and the lower it is, the better the model is trained.
Therefore, it is a suitable candidate for the \emph{\mbox{application-aware} figure of merit} in this exemplary application.
To minimize the KL~divergence, the model parameters are adapted via the covariance matrix adaption evolutionary strategy (\mbox{CMA-ES})~\cite{hansen2019pycma}, a \mbox{gradient-free} optimizer. 
\end{example}

\subsection{Compilation Environment}
A compilation environment is necessary that actually supports customizable figures of merit. 
Although this is not the case for the currently frequently used SDKs such as Qiskit and TKET, other tools offer this freedom, such as, e.g., the \emph{MQT~Predictor} framework~\cite{quetschlich2023mqtpredictor}.
This compilation framework allows one to use a customizable figure of merit and learns, based on artificial intelligence, the optimal sequence of compilation passes for the considered application.
To this end, \emph{reinforcement learning} is utilized to determine the most promising sequence of synthesis, optimization, and mapping passes from a large number of available passes provided by multiple quantum SDKs, such as Qiskit~\cite{qiskit} and TKET~\cite{tket}.

To realize the proposed approach, significant changes to the \emph{MQT~Predictor} framework were necessary. 
Previously, the framework would try to learn an optimized sequence of passes to compile arbitrary circuits to a particular device---following a \enquote{\mbox{one-fits-all}} approach similar to other SDKs.
For the use case considered here, only a single input circuit (aiming to solve a particular application instance) is considered and the goal is to find the compiled circuit that is expected to achieve the best application result.
Due to the variational nature of most applications currently being explored using quantum computing, the framework also had to be extended to handle the compilation of parameterized quantum circuits---a seemingly simple, yet daunting task in practice due to many subtleties across different quantum compilers and their handling of these circuits.
Last, but certainly not least, the framework had to be extended with the possibility to execute/simulate the resulting circuits in order to evaluate the \mbox{application-aware} figure of merit as well as running the application.

\begin{example}
To compile the underlying quantum circuit of the QCBM application for the KL~divergence as its \mbox{application-aware} figure of merit, the MQT~Predictor framework~\cite{quetschlich2023mqtpredictor} has been adapted accordingly.
By providing the quantum circuit of the considered application, such as, e.g., an application instance with $4$ qubits as shown in \autoref{fig:application}, the framework over time learns the most efficient combination of compilation passes that leads to the highest solution quality---while, during the training, the application itself is run for each determined sequence of compilation passes.
\end{example}

\section{Obtained Results}\label{sec:eval}
In this section, the application described in the running examples of \autoref{sec:proposed} is evaluated with three different problem sizes of $4$, $6$, and $8$ qubits using the described setup based on the MQT~Predictor framework~$v2.0.0$ and Qiskit~$v0.45.3$ in Python\footnote{The source code to create those results is \mbox{publicly-available} on GitHub (\url{https://github.com/cda-tum/mqt-predictor/tree/quantum_generative_modeling}).}.
For circuit execution, noisy simulators are used instead of actual devices due to their limited access and associated costs, although the proposed methodology is not limited to classical simulations and also works on actual devices\footnote{In fact, running on real devices can be expected to yield the most accurate and optimized results as it does not involve any modeling.}.
To that end, \mbox{mock-ups} of the \emph{ibmq\_quito}, \emph{ibmq\_nairobi}, and \emph{ibmq\_montreal} devices with $5$, $7$, and $27$ qubits are used, and each problem instance has been run on the smallest fitting device.
The results are shown in \autoref{fig:results}.

To evaluate the proposed \mbox{application-aware} compilation scheme, the solution quality of the considered application---in this case, the KL~divergence for the QCBM application---is used as a figure of merit with lower values representing better solutions.
Since the application itself runs for a certain number of epochs to optimize the parameter values of the underlying quantum circuit, the figure of merit generally decreases over the epochs, and the lowest KL~divergence reached at \emph{any} epoch constitutes the best solution.
To this end, the same compiled and parameterized quantum circuit is used throughout all episodes---just with different parameter values.
The proposed approach is compared to Qiskit's \mbox{most-optimized} compilation scheme~(O3) as the baseline to beat, as this is what was originally used for the application considered.
Furthermore, the results of Qiskit's default compilation scheme~(O1) are also denoted to underline the effect compilation has on the solution quality.
For both baselines, $25$ compilation runs have been carried out to cover the spread of their compilation outcomes and, therefore, solution quality.
To this end, the worst, median, and best run have been determined---again, measured by the lowest KL~divergence that occurred during those runs.

\def\reswidth{0.32}
\begin{figure*}[t]
    \centering
        \begin{subfigure}[t]{0.6\linewidth}
        \centering\includegraphics[width=1.0\linewidth]{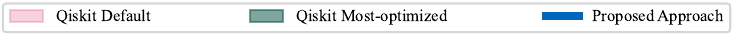}
    \end{subfigure}
    
    \begin{subfigure}[t]{\reswidth\linewidth}
        \centering\includegraphics[width=1.0\linewidth]{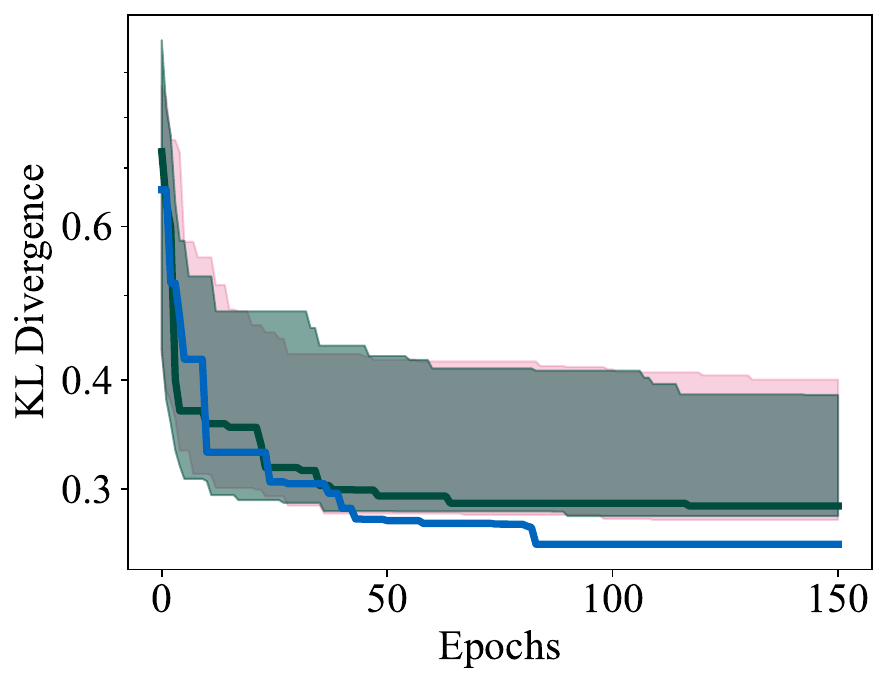}
        \caption{4 qubits on ibmq\_quito.}
        \label{fig:4qubits}
    \end{subfigure}
    \begin{subfigure}[t]{\reswidth\linewidth}
        \centering\includegraphics[width=1.0\linewidth]{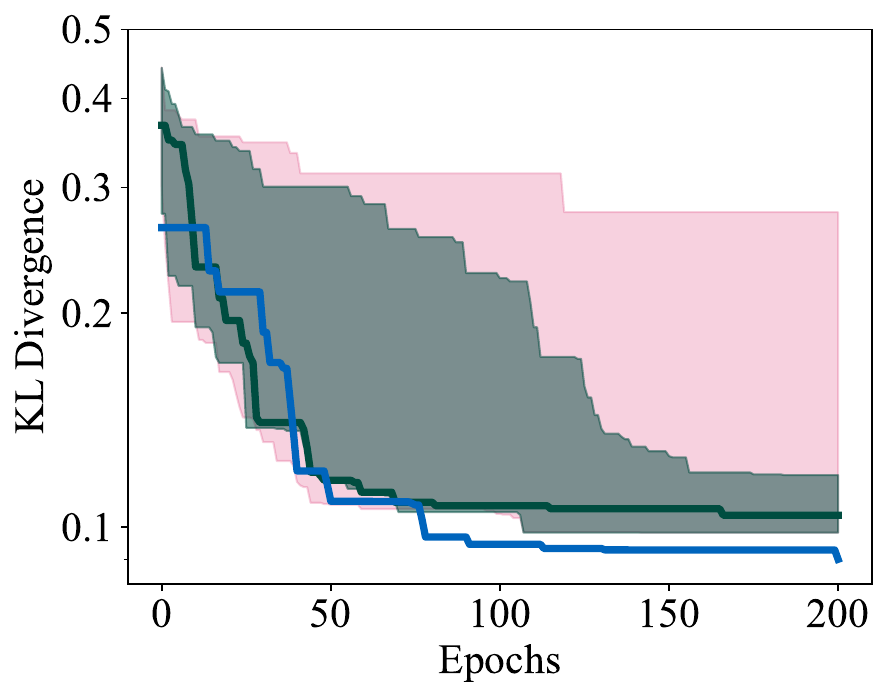}
        \caption{6 qubits on ibmq\_nairobi.}
        \label{fig:6qubits}
    \end{subfigure}
    \begin{subfigure}[t]{\reswidth\linewidth}
        \centering\includegraphics[width=1.0\linewidth]{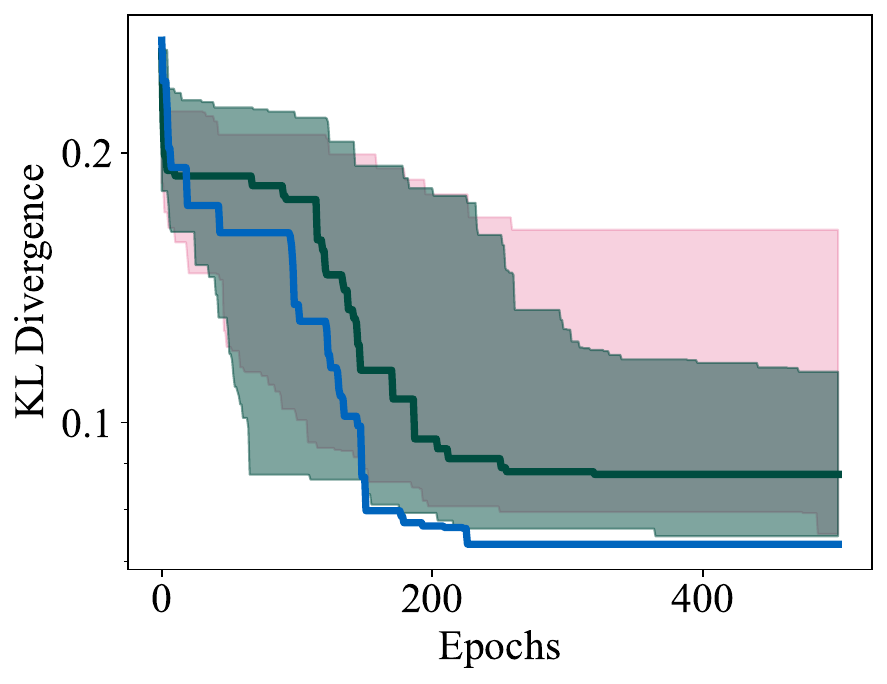}
        \caption{8 qubits on ibmq\_montreal.}
        \label{fig:8qubits}
    \end{subfigure}
    \caption{Results of the training process across three different circuit sizes. Lower values indicate a higher solution quality.}
    \label{fig:results}
\end{figure*}

The results for the $4$ qubits application instance (whose underlying quantum circuit has been depicted as the exemplary circuit shown in \autoref{fig:application}) executed on the \mbox{mock-up} of the \emph{ibmq\_quito} device are shown in \autoref{fig:4qubits}.
To visualize the solution quality of all $25$ runs of both the \mbox{most-optimized} and the default baselines, their spread is visualized as a tube graph between the best and worst solution quality of all $25$ runs at the respective epoch.
Additionally, the median compilation run of the \mbox{most-optimized} leading to the median solution quality is denoted within the corresponding tube.

The proposed approach surpasses both baselines, yielding the best KL~divergence, and hence, solution quality.
Meanwhile, both baselines show qualitatively similar convergence behavior and solution quality.
Measured by the minimal KL~divergence across all epochs and compared to the minimal KL~divergence of the best baseline, the proposed approach results in an improvement of 
\[
1-\frac{min_{proposed}}{{min_{best~baseline}}} \approx 6.21\%.
\]
This performance gain is achieved \emph{without any} adjustments of the encoding of the generative model as a quantum circuit and is achieved solely taking into account the actual application during the compilation.
Compared to the median/worst run of the \mbox{most-optimized} baseline, the improvement is even greater with $9.59\%/32.54\%$.
This performance improvement, even over the \mbox{most-optimized} baseline, highlights the potential of application-aware compilation.

For the application instance with $6$ qubits executed on the \mbox{mock-up} of the \emph{ibmq\_nairobi} device, the solution quality is shown in \autoref{fig:6qubits}.
Although the spread of the default baseline is significantly larger than that of the \mbox{most-optimized} one, its solution quality is comparable in the respective best cases.
Similarly to before, the proposed approach again results in the best solution quality with an improvement of $8.23\%/13.30\%/22.94\%$ compared to the best/median/worst run of the \mbox{most-optimized} baseline.
The $8$ qubit application instance that is executed on the \emph{ibmq\_montreal} device has led to qualitatively similar results, as shown in \autoref{fig:8qubits}.
Here, the proposed approach resulted in an improvement of $2.03\%/16.45\%/32.47\%$ compared to the best/median/worst run of the \mbox{most-optimized} baseline.
For the training itself, the number of necessary epochs to reach convergence increases with the application size due to the larger number of quantum circuit parameters.

Although the evaluations indicate a general trend of improving solution quality with larger application instances, this is mainly caused by the formulation of the underlying 2D~dataset.
The percentage of tiles that are used to represent the X letter (orange tiles in~\autoref{fig:application}) decreases with the application size.
For the shown $4$~qubit instance it is $8$ out of $16$ tiles, while for the $6$~qubit instance it is significantly lower with $12$ out of $36$ tiles.
To show how this relation affects the KL~divergence values, consider the case where the randomly initialized circuits yield the superposition---leading to an initial KL~divergence value of $0.69$ for the $4$ qubit instance and $0.29$ for the $6$ qubit instance.
Therefore, the relative differences in solution quality yielded by different compilation schemes for the same application instance are more meaningful than the absolute values reached across different application sizes.

Furthermore, it should be noted that the spread of both baselines varies significantly between the application instances considered and, respectively, used \mbox{mock-ups} of the actual devices during noisy simulation.
This is due to a combination of reasons: the calibration data of the devices being considered are rather different, the circuit width and depth are also different, and, lastly, there is randomness in the classical \mbox{CMA-ES} optimizer, too.
Therefore, more studies are needed to clearly determine the influence of each of these factors.

\section{Discussion}\label{sec:discussion}
Determining useful solutions for the application is a challenge in and of itself in the current NISQ era.
Therefore, no opportunity should be left unexplored to further increase the solution quality and \mbox{application-aware} compilation could be a further step towards meaningful use of quantum computing.

The proposed approach allows end users to further increase the quality of their solution without adjusting how they derive their quantum circuit.
However, it comes with a large effort in both setting up the proposed compilation scheme and running it---while the latter can be significantly reduced when executing on actual devices instead of using a noisy simulator.
Consequently, end users should determine whether this effort outweighs the gains in solution quality.

This work constitutes a first step towards exploring the direction of \mbox{application-aware} quantum compilation offering many possible paths to further explore this idea, e.g., when choosing between different quantum devices to run the considered application or even before selecting the most promising algorithm and encoding the problem, respectively.
Additionally, the influence of other factors such as the underlying noisy simulator or actual devices as well as the classical optimizers and their influence should be further investigated to fully explore the potential the proposed approach offers.

\section{Conclusion}\label{sec:conclusion}
Quantum circuit compilation greatly affects the quality of the solution when using quantum computers to realize a considered application and often makes the difference between a successful execution and obtaining completely random results.
Current \mbox{state-of-the-art} compilers offer a selectable optimization level as a \mbox{trade-off} between the compilation time and the \emph{quality} of the compiled circuit---based on \emph{figures of merit} aiming at resembling how successful a compiled circuit can be executed on a quantum computer and, by that, acting as a \emph{proxy} for the solution quality of the considered application.
However, they treat the \mbox{to-be-compiled} quantum circuit as a sequence of instructions without considering the actual application it realizes---wasting a yet untapped potential to increase the solution quality.
In this work, a different compilation approach has been proposed that optimizes directly for the solution quality \emph{itself} instead for any proxy of it.
To showcase the effectiveness over the current compilation schemes, the experimental evaluations for an industry-inspired quantum generative model showed an increased solution quality for all evaluated application instances compared to Qiskit's \mbox{most-optimized} compilation scheme.
By this, this work presents a first step towards \emph{application-aware} compilation.

 \section*{Acknowledgments}
 This work received funding from the European Research Council (ERC) under the European Union’s Horizon 2020 research and innovation program (grant agreement No. 101001318), was part of the Munich Quantum Valley, which is supported by the Bavarian state government with funds from the Hightech Agenda Bayern Plus, and has been supported by the BMWK on the basis of a decision by the German Bundestag through project QuaST, as well as by the BMK, BMDW, the State of Upper Austria in the frame of the COMET program, and the QuantumReady project within Quantum Austria (managed by the FFG).
AL and CAR were partly funded by the Bavarian State Ministry of Economic Affairs in the project BenchQC under Grant DIK-0425/03.

\printbibliography

\end{document}